# Molecular dynamics without molecules: searching the conformational space of proteins with generative neural networks

Gregory Schwing[1,5,‡], Luigi L. Palese[2,‡], Ariel Fernández[3,4], Loren Schwiebert[1], Domenico L. Gatti[5]

[1]Department of Computer Science, College of Engineering, Wayne State University, Detroit, USA
[2]Department of Basic Medical Sciences, Neurosciences and Sense Organs, Univ. of Bari Aldo Moro, Bari, Italy
[3]National Research Council (CONICET), Buenos Aires 1033, INIQUISUR-CONICET, Av. Alem 1253, Bahía Blanca 8000, Argentina
[4]Daruma Institute for Applied Intelligence, AF Innovation, Pharma Consultancy, 1005 Oakhurst Avenue, NC 27262, USA
[5]Department of Biochemistry, Microbiology, and Immunology, Wayne State University School of Medicine, Detroit, USA

[‡]These authors contributed equally to the study.

gregory.schwing@med.wayne.edu
luigileonardo.palese@uniba.it
ariel@afinnovation.com
loren@wayne.edu
dgatti@med.wayne.edu

## Abstract

All-atom and coarse-grained molecular dynamics are two widely used computational tools to study the conformational states of proteins. Yet, these two simulation methods suffer from the fact that without access to supercomputing resources, the time and length scales at which these states become detectable are difficult to achieve. One alternative to such methods is based on encoding the atomistic trajectory of molecular dynamics as a "shorthand" version devoid of physical particles, and then learning to propagate the encoded trajectory through the use of artificial intelligence. Here we show that a simple 'textual' representation of the frames of molecular dynamics trajectories as vectors of Ramachandran basin classes retains most of the structural information of the full atomistic representation of a protein in each frame, and can be used to generate equivalent atom-less trajectories suitable to train different types of generative neural networks. In turn, the trained generative models can be used to extend indefinitely the atom-less dynamics or to sample the conformational space of proteins from their representation in the models latent space. We define intuitively this methodology as *molecular dynamics without molecules*, and show that it enables to cover physically relevant states of proteins that are difficult to access with traditional molecular dynamics.

## Introduction

Molecular dynamics (MD) is a widely used tool to investigate molecular processes such as protein folding, protein conformational changes, and protein-ligand interactions (reviewed in [1]). In atomistic MD (all-atom MD, AAMD), the system consists of a collection of interacting particles describing both solute and solvent placed inside a sufficiently large simulation box, whose movements are described by Newton's laws of motions. An algorithm (i.e., leap-frog [2]) is employed to advance the system as a function of time over the course of many time steps. Typically, in AAMD, the system size is in the order of tens of thousands of atoms and the time step is 1–2 fs. To advance the state, the equations of motion are integrated stepwise, with forces acting on the system particles being computed from a *force field* that includes both *non-bonded* interactions and various types of *bonded* potentials. A large number of software packages are available for MD simulations of biomolecules (i.e., [3-6]).

Even with access to super-computing resources, the limited length (nm) and time scales (ns to μs) attainable in AAMD still pose a significant limitation to the study of molecular processes that occur on longer time scales (e.g., protein folding/unfolding, some protein conformational transitions) or larger length scales (e.g., protein complexes or protein–membrane systems). In standard AAMD, atoms serve as particles. In coarse-grained MD (CGMD), several atoms (both in the protein and solvent) are grouped together to form *beads* that serve as the system particles [7-10], so that the number of particles is significantly reduced in comparison to an AAMD representation of the same system. As a result, longer time and larger length scales are attainable by CGMD at the expense of atomistic details.

A different type of coarse-grained system for protein-folding dynamics was introduced by Fernández [11-13] based on a discretized representation of the conformational space using the basins of the Ramachandran map. In its basic form, the modulo-basin representation of the protein backbone torsional state is simply a text vector with an alphabet of 4 characters (i.e., 0,1,2,3), each character representing a basin of the Ramachandran map. Most recently, Fernández has shown that a deep learning neural network can be trained on the *textual* representation of an MD trajectory [14], and then used to propagate indefinitely the basins encoded dynamics. In this context, protein trajectories are defined by the evolution of the backbone torsional coordinates $[\phi_n, \psi_n]$ $n = 1,\ldots,N$ ($N$ = number of residues -2) represented as text vectors, and the conformation space, described mathematically as the so-called *modulo-basin quotient space* [15], becomes the Cartesian products of the Ramachandran basins (see Figure 1 in [14]). Since modulo-basin dynamics are effectively the particle-less equivalent of traditional AAMD and CGMD, we like to describe them intuitively as *molecular dynamics without molecules*. In this type of dynamics the transition timescale for each iteration is the minimal Ramachandran intra-basin equilibration time $\tau = 100$ ps [15], considerably larger than the fs integration timescales used in MD. For this reason, the propagated modulo-basin dynamics is more likely to capture infrequent inter-basin transitions that are hard to observe in traditional AAMD or CGMD.

Here we show that three types of generative neural networks, a sequence-to-sequence Transformer [16], a Variational Auto Encoder (VAE) [17, 18], and a Generative Adversarial Network (GAN) [19], trained on the basin-encoded *textual* representation of an ensemble of 200 MD trajectories displaying the open/closed conformational transition of the *Escherichia coli* enzyme Adenylate Kinase (AdK) [20], are capable of generating a wide range of protein backbone torsional states of AdK that are not observed in the training ensemble.

## Results

**Basin-encoded textual representation of MD trajectories.** The AdK enzyme featured in the training ensemble of MD trajectories consists of 214 residues, and thus only 212 pairs of $\phi$, $\psi$ dihedral angles (excluding first and last residue) can be measured for each time step of a trajectory. **Figure 1A** show a Ramachandran plot of the collected torsional data from the first of the ensemble of MD trajectories used in this work (here onward referred to as the AdK set), which contained 97 frames. Each point in the plot represents a backbone torsional state of one residue in one of the frames of the trajectory.

Using a *KMeans* algorithm, points in the plot were classified as belonging to 1 of 4 clusters (basins) labeled 0, 1, 2, 3 respectively (**Figure 1B**). The centroids of each cluster were recorded and stored for later use in the reconstruction of a full atomistic representation of AdK from the basin encoded torsional states. Each frame of the trajectory was then converted to a character vector with an alphabet of 4 symbols corresponding to the 4 classes/clusters. As an example, the basin-encoded *textual* representation of three distinct frames in the trajectory is shown in **Figure 2**. A similar encoding process was carried for all 200 trajectories in the AdK set.

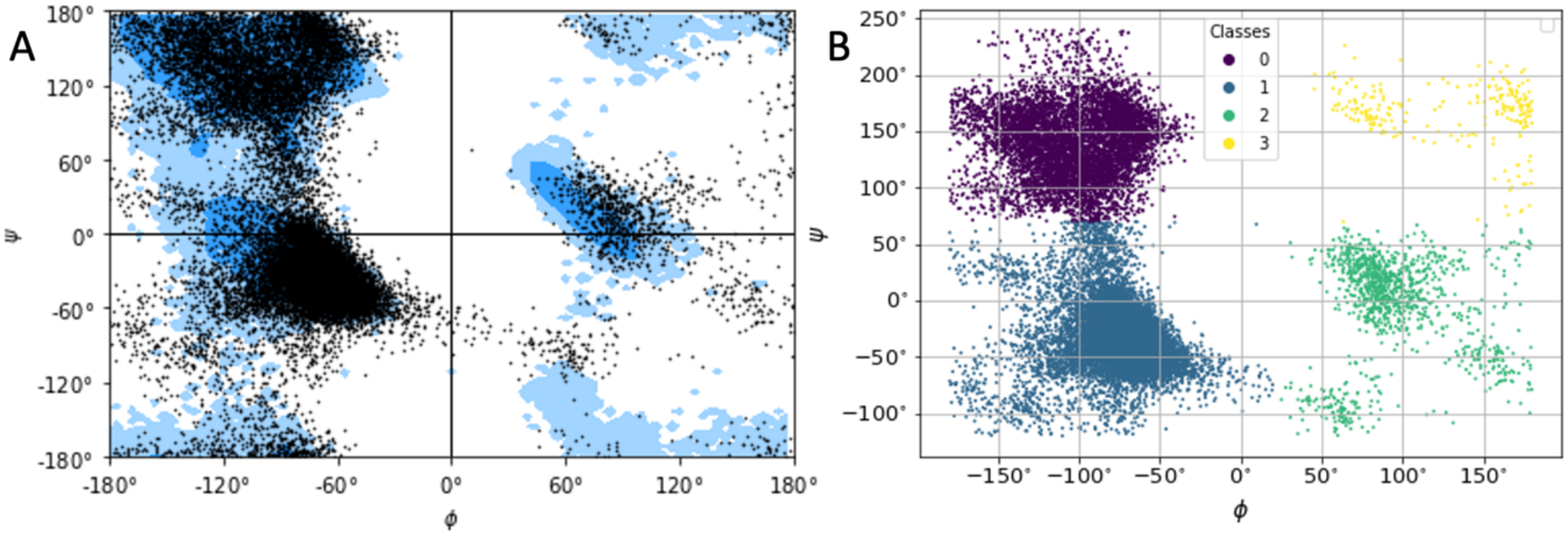

**Figure 1. A.** Ramachandran plot of the first trajectory in the AdK set. Each point in the plot is a $\phi$, $\psi$ pair of one residue in one frame of the trajectory. **B.** Classification of the points in the Ramachandran plot as belonging to 1 of 4 basins.

Frame 1

```
0 0 0 0 0 3 0 0 2 1 2 1 1 1 1 1 1 1 1 1 1 1 1 2 0 0 0 0 0 1 1 1 1 1 1 1 1 1 1 1 2 0 1 0 2 1
1 1 1 1 1 1 1 1 2 0 0 0 0 1 1 1 1 1 1 1 1 1 1 1 1 1 0 1 1 1 1 1 3 0 0 0 0 2 0 0 1 0 1 1 1 1
1 1 1 1 1 1 2 0 0 0 1 0 0 0 0 0 0 0 0 1 1 1 1 1 1 1 1 1 1 0 0 0 0 1 1 1 2 0 0 0 0 1 1 1 0 0
0 1 0 0 2 0 0 1 1 1 2 0 0 0 0 0 0 1 1 1 0 1 1 1 1 1 1 1 1 1 1 1 1 1 1 1 1 1 1 1 1 1 1 1 1 1
1 1 1 2 1 0 0 0 0 0 0 0 1 1 0 0 1 1 1 1 1 1 1 1 1 1 1 1
```

Frame 50

```
0 0 0 0 0 3 0 1 1 1 0 0 2 1 1 1 1 1 1 1 1 1 1 2 0 0 0 0 0 0 2 1 1 1 1 1 1 1 1 1 2 0 1 0 2 1
1 1 1 1 1 1 1 1 1 0 1 0 0 1 1 1 1 1 1 1 1 1 1 1 1 1 0 1 1 1 1 1 0 0 0 0 0 2 0 0 1 0 1 1 1 1
1 1 1 1 1 1 2 1 0 0 1 0 0 0 0 0 0 0 0 1 1 1 1 1 1 1 1 1 1 0 0 0 0 1 1 1 2 0 0 0 3 1 1 1 0 0
0 1 0 0 2 1 0 1 1 1 2 0 0 0 0 0 0 1 1 0 0 1 1 1 1 1 1 1 1 1 1 1 1 1 1 1 1 1 1 1 1 1 1 1 1 1
1 1 1 2 1 0 0 0 0 0 0 0 1 1 0 0 1 1 1 1 1 1 1 1 1 1 1 1
```

Frame 97

```
0 0 0 0 0 3 0 1 2 1 1 0 2 1 1 1 1 1 1 1 1 1 1 2 0 0 0 0 0 1 1 1 1 1 1 1 1 1 1 1 0 0 1 1 1 1
1 1 1 1 1 1 1 1 2 0 0 0 0 1 1 1 1 1 1 1 1 1 1 1 1 1 0 1 1 1 1 1 3 0 0 0 0 2 0 0 1 0 1 1 1 1
1 1 1 1 1 1 2 1 0 0 1 0 0 0 0 0 0 0 0 1 1 1 1 1 1 1 1 1 1 0 0 0 0 1 1 1 2 0 0 0 0 1 1 1 0 0
0 1 0 0 2 0 0 1 1 1 2 0 0 0 0 0 0 1 1 1 1 1 1 1 1 1 1 1 1 1 1 1 1 1 1 1 1 1 1 1 1 1 1 1 1 1
1 1 1 2 1 0 0 0 0 0 0 0 1 1 0 0 1 1 1 1 1 1 1 1 1 1 1 1
```

**Figure 2.** Basin encoded textual representation of three frames in the first trajectory of the AdK set.

**Reconstruction of a full atomistic representation of AdK from a basin encoded representation of backbone angles.** Each basin-encoded frame was converted to a vector of $\phi$, $\psi$ angle pairs assigning to each character the $\phi$, $\psi$ values of the corresponding cluster centroid. As an example, conversion of the first 10 characters of frame 97 in Figure 2 to $\phi$, $\psi$ angle pairs (in radians) is shown in **Figure 3A**. The *Modeller* software package was then used to energy minimize a reference AdK structure imposing as constraints the $\phi$, $\psi$ angle values derived from the centroids. The Ramachandran plot of a reference AdK structure (in this case, frame 97 of the first trajectory in the AdK set) is shown in **Figure 3B**. The Ramachandran plot of the energy minimized structure with centroid derived $\phi$, $\psi$ constraints is shown in **Figure 3C**. Superimposed reference and $\phi$, $\psi$ centroid constraints minimized AdK structures (backbone RMSD: 1.213 Å) are shown

as a cartoon in **Figure 3D**. The overall conformation of the reference structure is preserved in the constraints minimized structure, despite individual $\phi$, $\psi$ angles pairs being set, based on the basin encoding of each frame, at the values of the corresponding $\phi$, $\psi$ centroids for the entire trajectory.

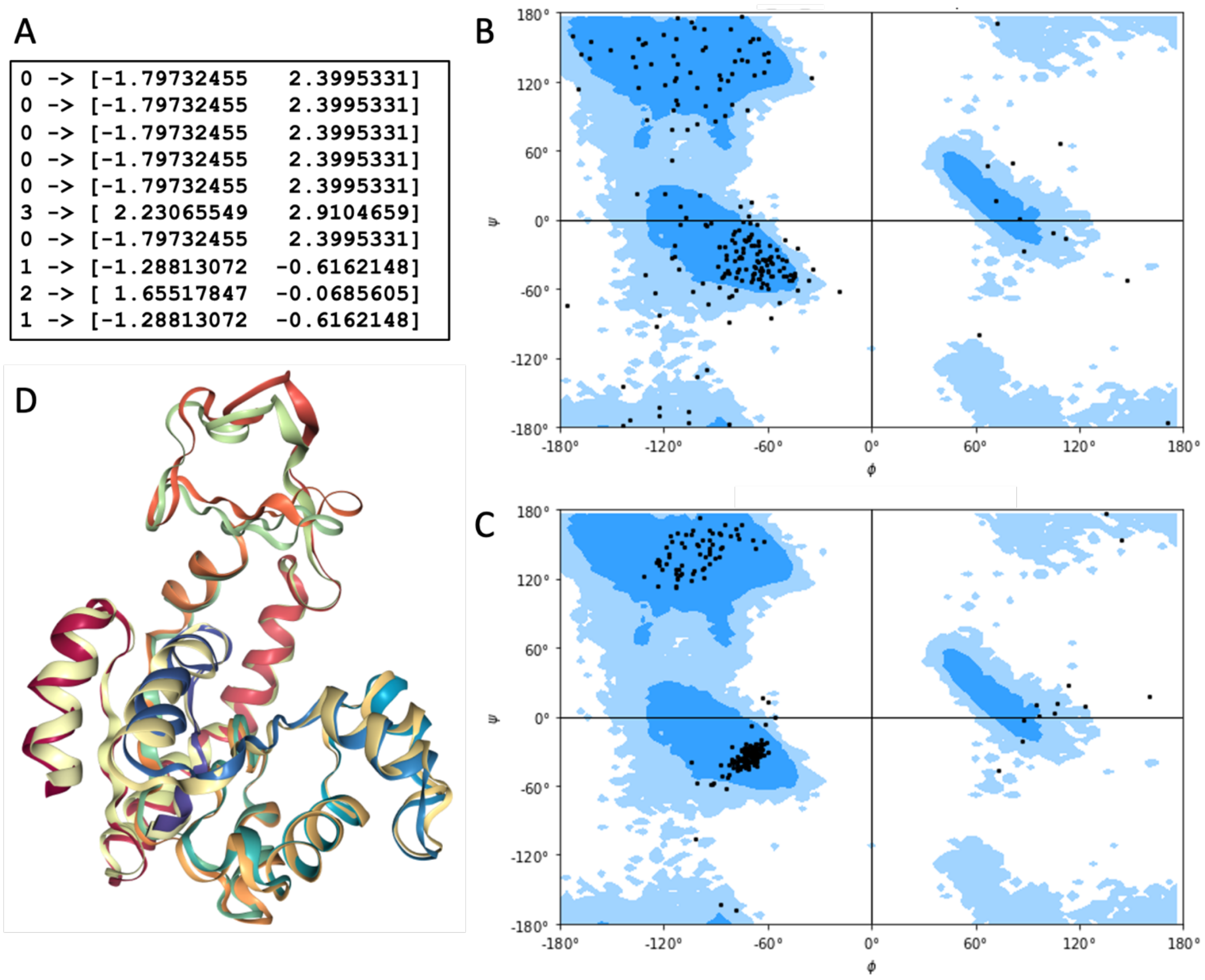

**Figure 3. A.** Conversion of the first 10 characters of frame 97 (**Figure 2**) to centroid $\phi$, $\psi$ angle pairs (in radians). **B.** Ramachandran plot of the reference AdK structure. **C.** Ramachandran plot of the reference AdK structure energy minimized with centroid derived $\phi$, $\psi$ constraints. **D.** Superposition of the reference and energy minimized AdK structures.

**Sequence-to-sequence Transformer.** A sequence-to-sequence model takes a sequence as input and translates it into a different sequence (i.e., the English to Spanish translation of a phrase). This is the main task performed in most applications of *natural language processing*. In our case we are interested in converting basin-encoded frame *n* (the *source*) into basin-encoded frame *n+1* (the *target*). Each element of the sequence can be used as such or further represented as some type of *token*.

A sequence-to-sequence model operates differently during *training* and *inference*. During training, an *encoder* turns the *source* into an intermediate representation. A *decoder* learns to predict the next *token i* in the *target* by looking at all previous *tokens* (*0:i-1*) in the *target* and at the *encoder* representation of the *source*.

During inference the *target* is predicted one token at a time. To this end, the *decoder* looks at the encoded *source* and at an initial "start" token (i.e., an initial padded character 4 in a basin-encoded frame), and uses them to predict the first real token in the *target* sequence. The predicted sequence is fed back to the *decoder*, to generate the next token, until it generates a "stop" token (i.e., a final padded character 5 in a basin encoded frame).

The use of a *frame-to-frame* Transformer in the extension of basin-encoded trajectories was recently advocated by Fernández [21], based on the important role played in a Transformer model by *neural attention* [16]. The Transformer *encoder* uses self-attention to produce a context-aware representations of each token in the input sequence. In the case of a basin-encoded frame this means that each basin in the basin-encoded sequence is assigned some *relevance* with respect to all other basins in the sequence. The Transformer *decoder* reads tokens 0...*N* in the target sequence and uses neural attention to identify which tokens in the encoded source sequence are most relevant to the target token *N*+1 that is trying to predict. Since self-attention is a process focused on the relationships between pairs of sequence elements, but is otherwise agnostic to the position of these elements in the sequence, information on the reciprocal influence of basins at different positions in a basin-encoded frame (that is, the cooperative contribution to the protein fold of residues of different types and positions in sequence space) was introduced in our Transformer model in the form of *Positional Embedding* layers. Two types of positional embeddings were implemented in these layers, relating the basins in a basin-encoded frame to their numbered positions in the protein sequence (*via* a vector of consecutive integers up to the sequence length), and to the type of amino acid at each position (*via* a vector of the same length built from an alphabet of 20 integers).

Starting from an initial basin-encoded dataset, training the *frame-to-frame* Transformer required the generation of three distinct datasets each with different padding:

1. An input dataset for the encoder with final padding [6 6 6].
2. An input dataset for the decoder with initial padding [4] (where 4 is the 'sequence start' signal) and final padding [5 6] (where 5 is the 'sequence end' signal).
3. A target dataset for the Transformer with final padding [5 6 6].

The encoder input (Train 1 input layer in **Table 1**) is the last know frame. The decoder input (Train 2 input layer in **Table 1**) is the same sequence as the transformer target (the next frame to predict). However they are padded differently, as the decoder input must start from a generic 'sequence start' signal (in this case, number 4) to predict the 1st basin of the target, and stop predicting when it encounters a 'sequence end' signal (in this case, number 5).

**Table 1.** Architecture of the *frame-to-frame* Transformer

| Layer (type) | Output Shape | Param # | Connected to |
|---|---|---|---|
| train1 (InputLayer) | [(None, None)] | 0 | [] |
| train2 (InputLayer) | [(None, None)] | 0 | [] |
| EncoderEmbedding (PosEmbedding) | (None, 215, 32) | 7872 | ['train1[0][0]'] |
| DecoderEmbedding (PosEmbedding) | (None, 215, 32) | 7872 | ['train2[0][0]'] |
| Encoder (TransformerEncoder) | (None, 215, 32) | 21120 | ['EncoderEmbedding[0][0]'] |
| Decoder (TransformerDecoder) | (None, 215, 32) | 37984 | ['DecoderEmbedding[0][0]', 'Encoder[0][0]'] |
| Dropout (Dropout) | (None, 215, 32) | 0 | ['Decoder[0][0]'] |
| Softmax (Dense) | (None, 215, 7) | 231 | ['Dropout[0][0]'] |

Total params: 75,079
Trainable params: 75,079
Non-trainable params: 0

The numbers of trainable parameters in each component of the Transformer model used in this study are shown in **Table 1**.

Trajectories 1:40 and trajectories 81:200 in the AdK set were used as the validation and training datasets, respectively, after conversion to basin-encoded frames. Accuracy and loss achieved by the Transformer model of AdK with the training and validation sets of basin-encoded trajectories are shown in **Figure 4**.

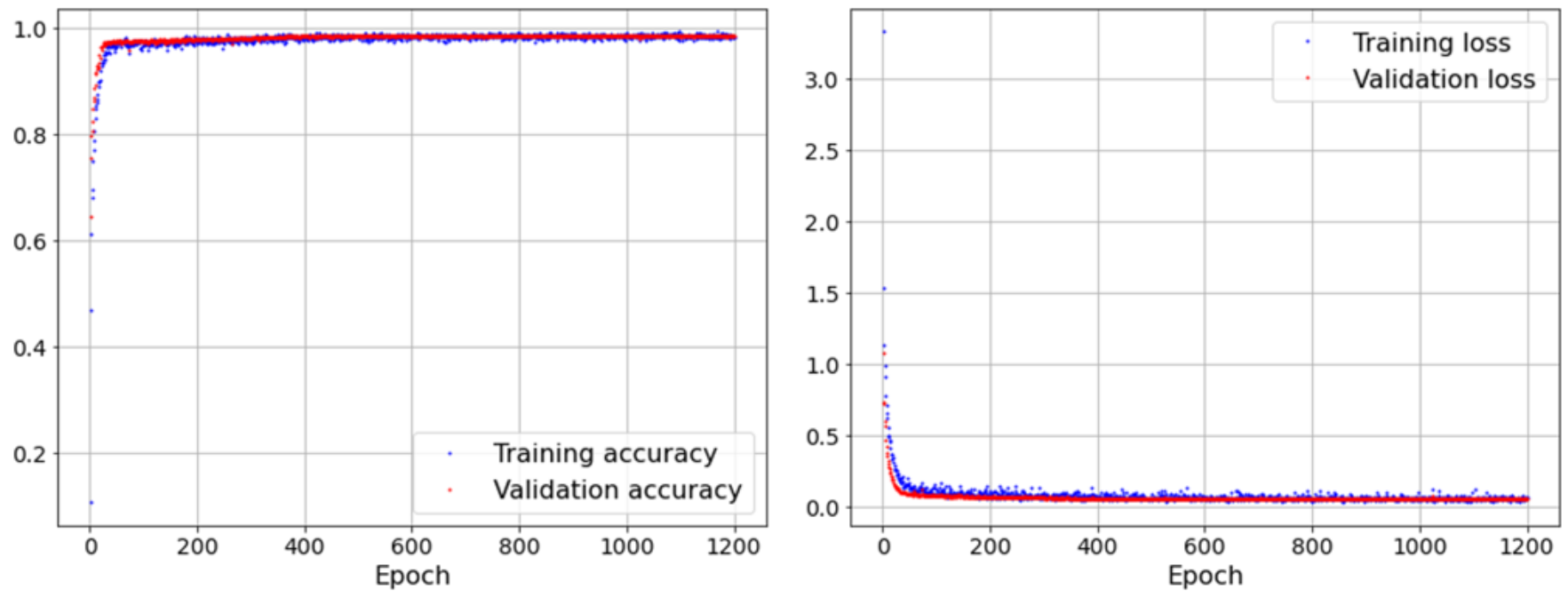

**Figure 4.** Training and validation set accuracy and loss for the *frame-to-frame* Transformer.

The trained *frame-to-frame* Transformer model was used to generate 100 basin encoded-frames using as input the last frame in the 1st trajectory of the validation set. The conformation of AdK in this frame is shown in the boxed panel on the top left of **Figure 5**. Snapshots from the Transformer derived modulo-basin extension of the validation trajectory, displaying a wider open state not observed in any of the trajectories of the AdK set, are shown in the other panels of **Figure 5**.

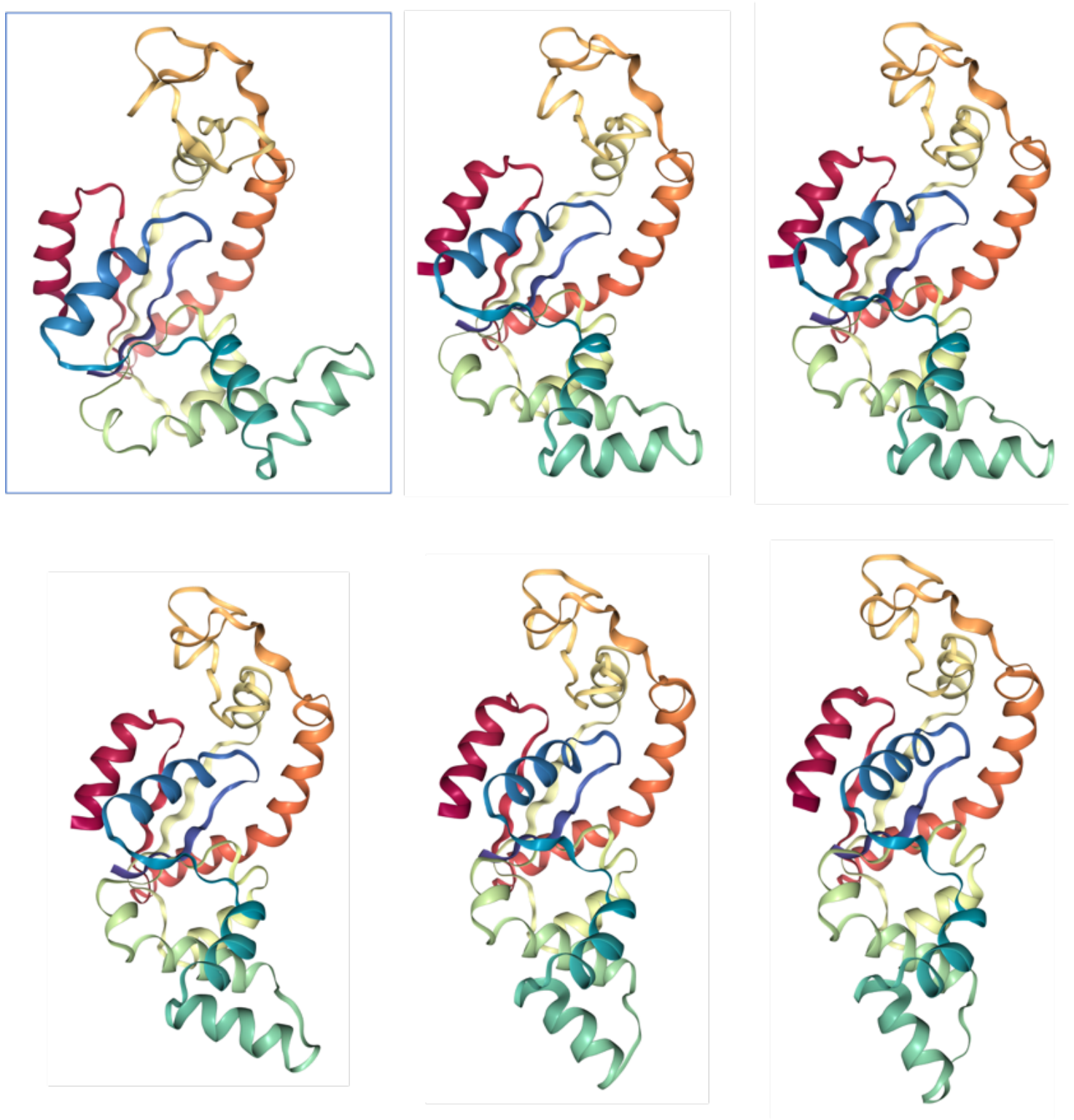

**Figure 5.** Transformer extended trajectory. **Boxed panel.** Open conformation of AdK observed in the 1st MD trajectory of the AdK set. **Other panels.** Snapshots of Modeller reconstructed conformations of AdK as derived from the extended basin-encoded trajectory.

**Variational Auto Encoder (VAE).** A typical variational autoencoder (VAE) turns an image into the parameters of a statistical distribution. The VAE then uses these parameters to randomly sample one element of the distribution, and decodes that element back to an image. Accordingly, the VAE used in this study features an *encoder* that turns an input basin-encoded frame (a vector) into two parameters (*z_mean* and *z_log_variance*) of a latent space normal distribution of trajectory frames representations. Upon randomly sampling a point *z* from the latent normal distribution, a *decoder* maps this point back to a basin-encoded trajectory frame. The process ensures that every point that is close to the latent space representation of a basin-encoded frame can be decoded to a similar vector of basins, thus forcing every direction in the latent space to be a meaningful axis of variation of backbone torsional coordinates.

Starting from an initial basin-encoded dataset, training the VAE required padding each basins vector of 212 characters with two initial and two final 0's (yielding a vector of 216 characters) in order to ensure proper downsizing with the 2-strided 1D-convolution layers of the *encoder* and upsizing with the 2-strided 1D-transposed convolution layers of the *decoder*. Samples and targets were the same vectors of basins. The numbers of trainable parameters in each component of the VAE used in this study are shown in **Table 2**.

The VAE parameters were trained *via* a reconstruction loss, forcing the decoded basin vectors to match the initial inputs, and a regularization loss, forcing the encoder output to be as close as possible to a 2D normal distribution centered around 0 at convergence. For the regularization loss, we used the Kullback–Leibler (KL) divergence. Reconstruction loss and KL loss for the training of a VAE model with the basin-encoded AdK set are shown in **Figure 6**.

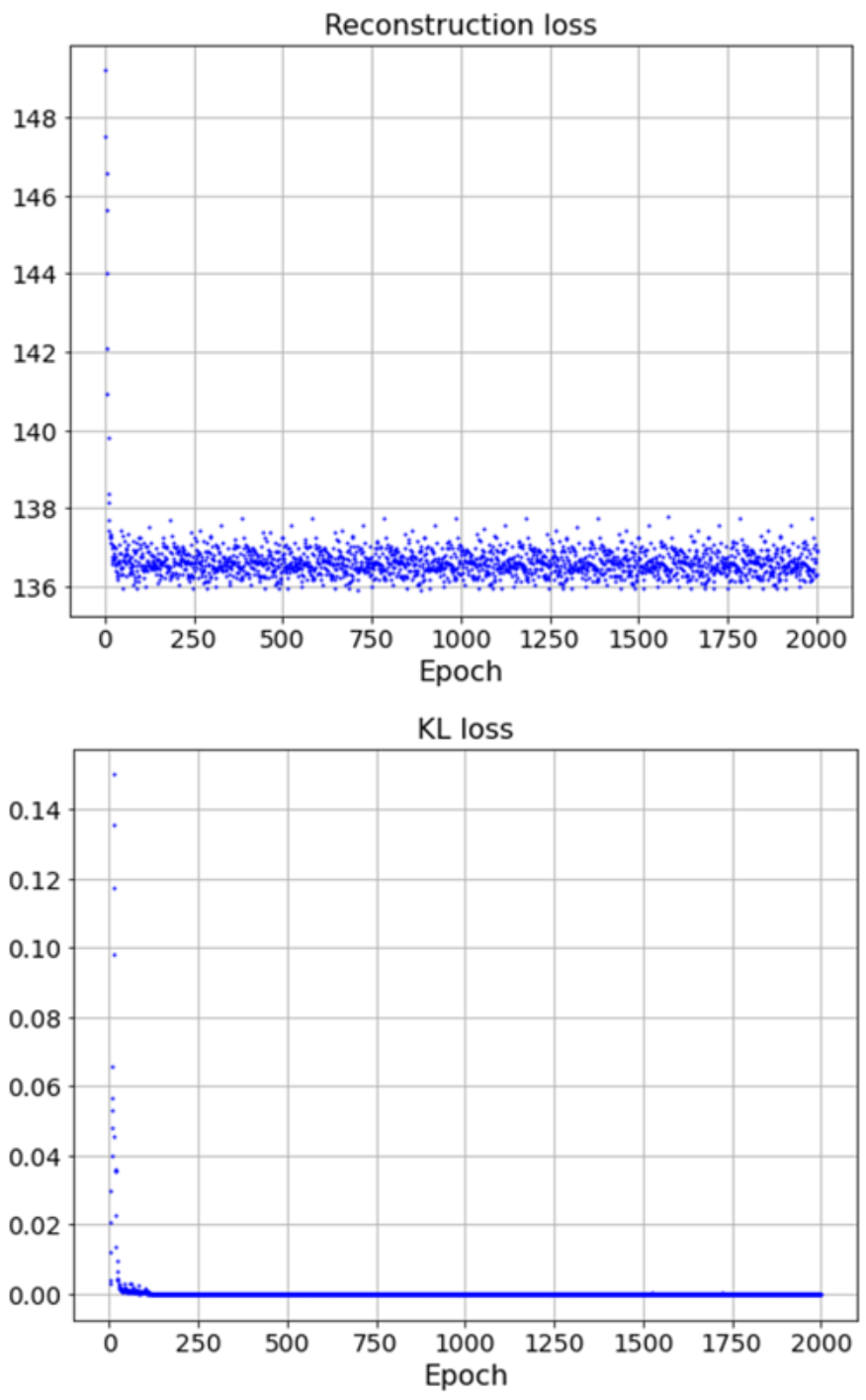

**Figure 6.** Reconstruction and *KL* loss for the training of a VAE model with the basin-encoded AdK set.

**Table 2.** Architecture of the VAE network used in this study.

Model: "encoder"

| Layer (type) | Output Shape | Param # | Connected to |
|---|---|---|---|
| input_1 (InputLayer) | [(None, 216, 1)] | 0 | [] |
| conv1d (Conv1D) | (None, 108, 32) | 160 | ['input_1[0][0]'] |
| conv1d_1 (Conv1D) | (None, 54, 64) | 8256 | ['conv1d[0][0]'] |
| flatten (Flatten) | (None, 3456) | 0 | ['conv1d_1[0][0]'] |
| dense (Dense) | (None, 16) | 55312 | ['flatten[0][0]'] |
| z_mean (Dense) | (None, 2) | 34 | ['dense[0][0]'] |
| z_log_var (Dense) | (None, 2) | 34 | ['dense[0][0]'] |

Total params: 63,796
Trainable params: 63,796
Non-trainable params: 0

Model: "decoder"

| Layer (type) | Output Shape | Param # |
|---|---|---|
| input_1 (InputLayer) | [(None, 2)] | 0 |
| dense (Dense) | (None, 3456) | 10368 |
| reshape (Reshape) | (None, 54, 64) | 0 |
| conv1d_transpose (Conv1DTranspose) | (None, 108, 64) | 16448 |
| conv1d_transpose_1 (Conv1DTranspose) | (None, 216, 32) | 8224 |
| conv1d (Conv1D) | (None, 216, 1) | 129 |

Total params: 35,169
Trainable params: 35,169
Non-trainable params: 0

A representation of the trained VAE latent space is shown in the top left panel of **Figure 7**. Evenly spaced points (red dots) along a trajectory vector in this latent space (red line) were used by the VAE *decoder* to reconstruct basins vectors representing chemically sensible torsional coordinates. The software package Modeller was again used to minimize a reference AdK structure (last frame in the 1$^{st}$ trajectory of the AdK set) imposing as constraints the $\phi$, $\psi$ angle values derived from the basins vectors. Snapshots from the trajectory in VAE latent space, displaying again a wider open state not observed in any of the trajectories of the AdK set, are shown in the other panels of **Figure 7**.

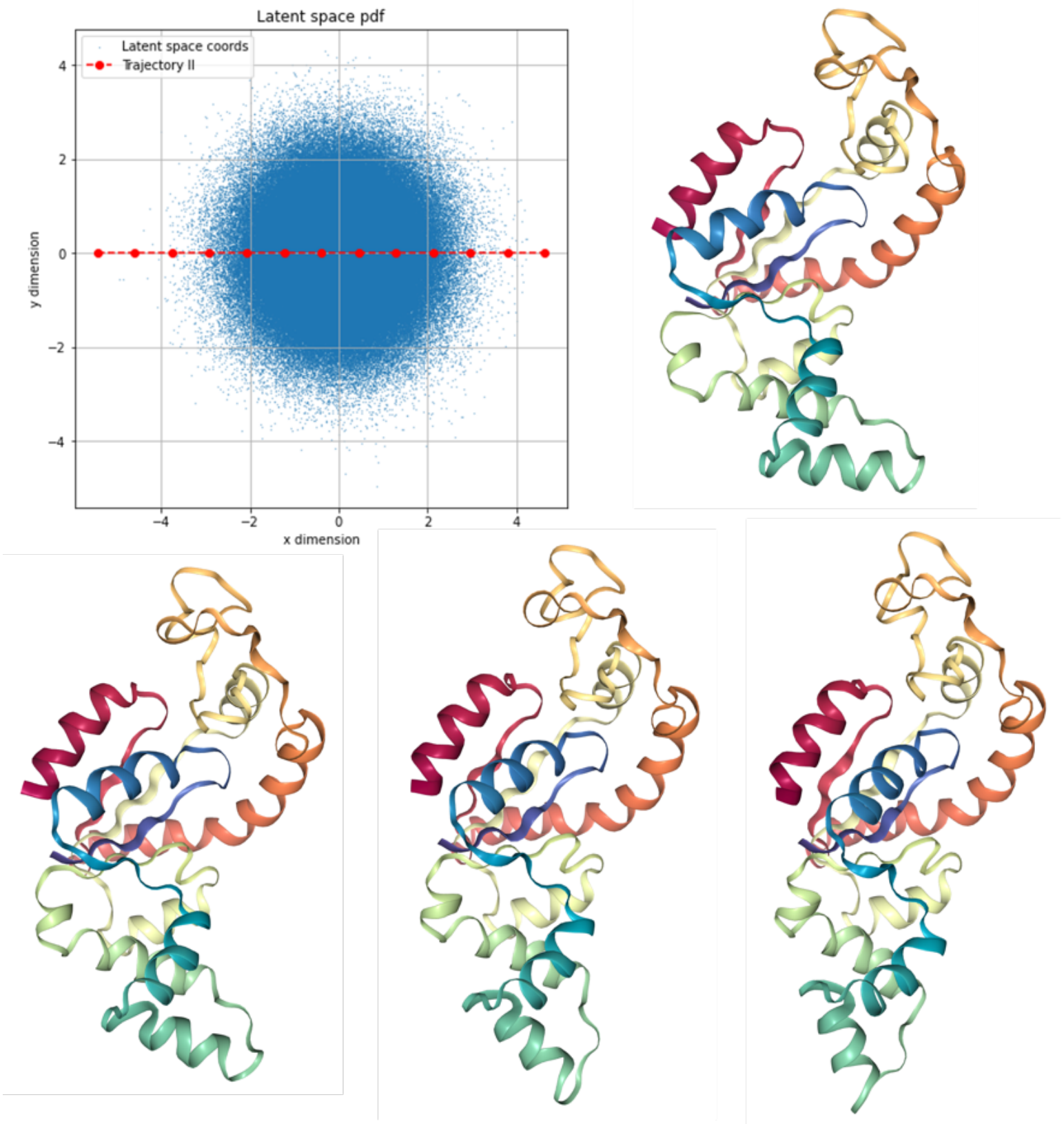

**Figure 7. Boxed panel.** Representation of the VAE latent space. Evenly spaced points (red dots) along a trajectory vector in this space were used by the VAE *decoder* to reconstruct basins vectors. **Other panels.** Snapshots from the trajectory in VAE latent space.

**Generative Adversarial Network (GAN).** A Generative adversarial network (GAN) is an alternative to a Variational Auto Encoder (VAE) for learning latent spaces of images. It enables the generation of images that are statistically indistinguishable from real ones. A typical GAN has two components: a *Generator* and a *Discriminator*. The *Generator* takes as input a random point in a latent space, and decodes it to a synthetic image. A *Discriminator* takes as input an image (real or synthetic), and predicts whether the image came from the training set or was created by the generator. As training goes on, the generator learns to produce increasingly realistic images, while the discriminator adapts to the improving capabilities of the generator, setting a progressively higher bar of realism for the generated images. In particular, in the GAN used in this study the generator maps random vectors of length 128 (the dimensions of the latent space) to vectors of length 216 (same as the VAE input) that look like basin-encoded frames of an MD trajectory, while the discriminator takes as input a candidate basins vector (real or synthetic) and classifies it as one of two classes: "generated basins vector" or "real basins vector that comes from the training set". The GAN network chains the generator and the discriminator together, mapping latent space vectors to the discriminator's classification as "fake" or "real" of the basins vectors that the generator decodes from the latent vectors. The discriminator is trained on examples of real and fake basins vectors paired with real and fake labels, just as a regular classification model. The generator is trained with targets that always say "these are real basins vectors." This updates the generator weights toward getting the discriminator to predict as "real" the generated basins vectors. Ultimately, the generator learns to fool the discriminator because at every step it becomes more likely for the discriminator to classify as real the basins vectors decoded by the generator. The numbers of trainable parameters in each component of the GAN used in this study are shown in **Table 3**.

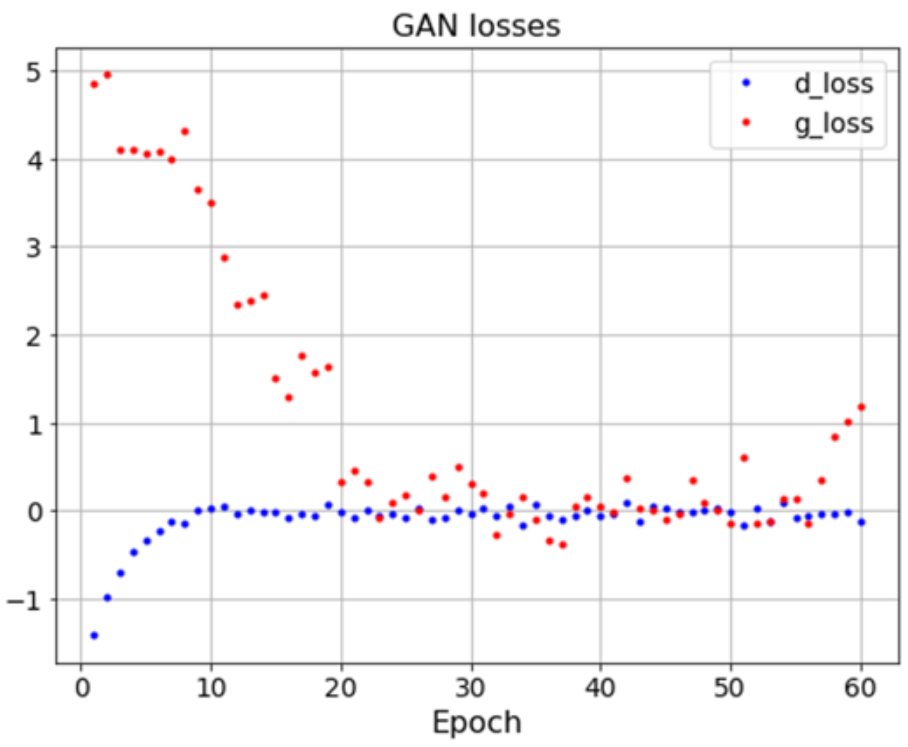

**Figure 8.** Discriminative and adversarial losses during the training of a GAN with the basin-encoded AdK set.

Training a GAN requires the use of separate optimizers for the generator and the discriminator, and the calculation of a loss for each optimizer (*g_loss* and *d_loss*, respectively). For this training, we adopted the Wasserstein GAN (WGAN) with Gradient Penalty (GP) implementation of the original Wasserstein GAN [22]. The code was modified from the Keras [23] WGAN_GP example written by A.K. Nain to work with basins vectors rather than images. In this type of GAN implementation, both *g_loss* and *d_loss* are expected to ultimately converge to 0. This was indeed observed during the training of our WGAN_GP with the basin-encoded AdK set (**Figure 8**).

**Table 3.** Architecture of the generator and discriminator modules of the GAN used in this study.

Model: "generator"

| Layer (type) | Output Shape | Param # |
|---|---|---|
| dense_1 (Dense) | (None, 1152) | 148608 |
| reshape (Reshape) | (None, 9, 128) | 0 |
| conv1d_transpose (Conv1DTranspose) | (None, 27, 128) | 81920 |
| batch_normalization (BatchNormalization) | (None, 27, 128) | 512 |
| leaky_re_lu_4 (LeakyReLU) | (None, 27, 128) | 0 |
| conv1d_transpose_1 (Conv1DTranspose) | (None, 54, 64) | 40960 |
| batch_normalization_1 (BatchNormalization) | (None, 54, 64) | 256 |
| leaky_re_lu_5 (LeakyReLU) | (None, 54, 64) | 0 |
| conv1d_transpose_2 (Conv1DTranspose) | (None, 108, 32) | 10240 |
| batch_normalization_2 (BatchNormalization) | (None, 108, 32) | 128 |
| leaky_re_lu_6 (LeakyReLU) | (None, 108, 32) | 0 |
| conv1d_transpose_3 (Conv1DTranspose) | (None, 216, 32) | 5120 |
| batch_normalization_3 (BatchNormalization) | (None, 216, 32) | 128 |
| leaky_re_lu_7 (LeakyReLU) | (None, 216, 32) | 0 |
| conv1d_4 (Conv1D) | (None, 216, 1) | 160 |
| batch_normalization_4 (BatchNormalization) | (None, 216, 1) | 4 |
| activation (Activation) | (None, 216, 1) | 0 |

Total params: 288,036
Trainable params: 287,522
Non-trainable params: 514

Model: "discriminator"

| Layer (type) | Output Shape | Param # |
|---|---|---|
| conv1d (Conv1D) | (None, 108, 16) | 96 |
| leaky_re_lu (LeakyReLU) | (None, 108, 16) | 0 |
| conv1d_1 (Conv1D) | (None, 54, 32) | 2592 |
| leaky_re_lu_1 (LeakyReLU) | (None, 54, 32) | 0 |
| dropout (Dropout) | (None, 54, 32) | 0 |
| conv1d_2 (Conv1D) | (None, 27, 64) | 10304 |
| leaky_re_lu_2 (LeakyReLU) | (None, 27, 64) | 0 |
| dropout_1 (Dropout) | (None, 27, 64) | 0 |
| conv1d_3 (Conv1D) | (None, 9, 128) | 41088 |
| leaky_re_lu_3 (LeakyReLU) | (None, 9, 128) | 0 |
| dropout_2 (Dropout) | (None, 9, 128) | 0 |
| flatten (Flatten) | (None, 1152) | 0 |
| dropout_3 (Dropout) | (None, 1152) | 0 |
| dense (Dense) | (None, 1) | 1153 |

Total params: 55,233
Trainable params: 55,233
Non-trainable params: 0

Differently from a VAE, the latent space of a GAN is not continuous, and thus it is not amenable to decode points along a specific trajectory vector (as shown in **Figure 7**). For this reason, the trained GAN generator was used to decode 200 random vectors of length 128 (the latent space dimensions) drawn from a standard normal distribution. This sampling of the GAN latent space is shown in the top left panel of **Figure 9** with the random vectors rotated at evenly spaced angles for ease of display. Snapshots of the AdK conformations reconstructed from basins vectors decoded by the generator from latent space points are shown in the other panels of **Figure 9**. GAN sampling of the conformational space of AdK produces the most dramatic divergence from the conformations observed in all the trajectories of the AdK set.

## Conclusions.

MD simulations have been for over two decades an essential tool to gain insight into the thermodynamic properties and detectable states of proteins [24, 25]. However, statistically significant sampling of proteins conformation space is difficult to achieve, and important but rare events can be easily missed in AAMD simulations. Significant advances in extending MD times and length scales have been achieved with coarse grained methods (CGMD). Yet these methods require a difficult joint optimization of the grouping of atoms into the beads that serve as the system particles, and of the force field used to calculate the interactions between these particles. In the face of these limitations, alternative approaches have been proposed to extend MD time and length scales to physically relevant quantities. One such approach is based on encoding the MD trajectory as a "shorthand" version in a *modulo-basin quotient space* made up of the Cartesian products of the Ramachandran basins [15], and then learning to propagate the encoded trajectory through the use of artificial intelligence [14]. Here we have shown that a simple 'textual' representation of the frames of MD trajectories as vectors of Ramachandran basin classes retains most of the structural information of the full atomistic representation of the protein in each frame, and can be used to generate an equivalent dataset of atom-less trajectories (these are effectively the *projection* of particle-based trajectories into modulo-basin quotient space) suitable to train different types of generative neural networks. In turn, the trained generative models can be used to extend indefinitely the atom-less dynamics (as is the case of the *frame-to-frame* Transformer model, **Figure 5**) or to sample the conformational space of proteins from their representation in a structured latent space (as is the case for the chemically meaningful trajectories that cross the continuous latent space of a VAE model (**Figure 7**), or from random points in the unstructured but more vast latent space of a GAN model (**Figure 9**).

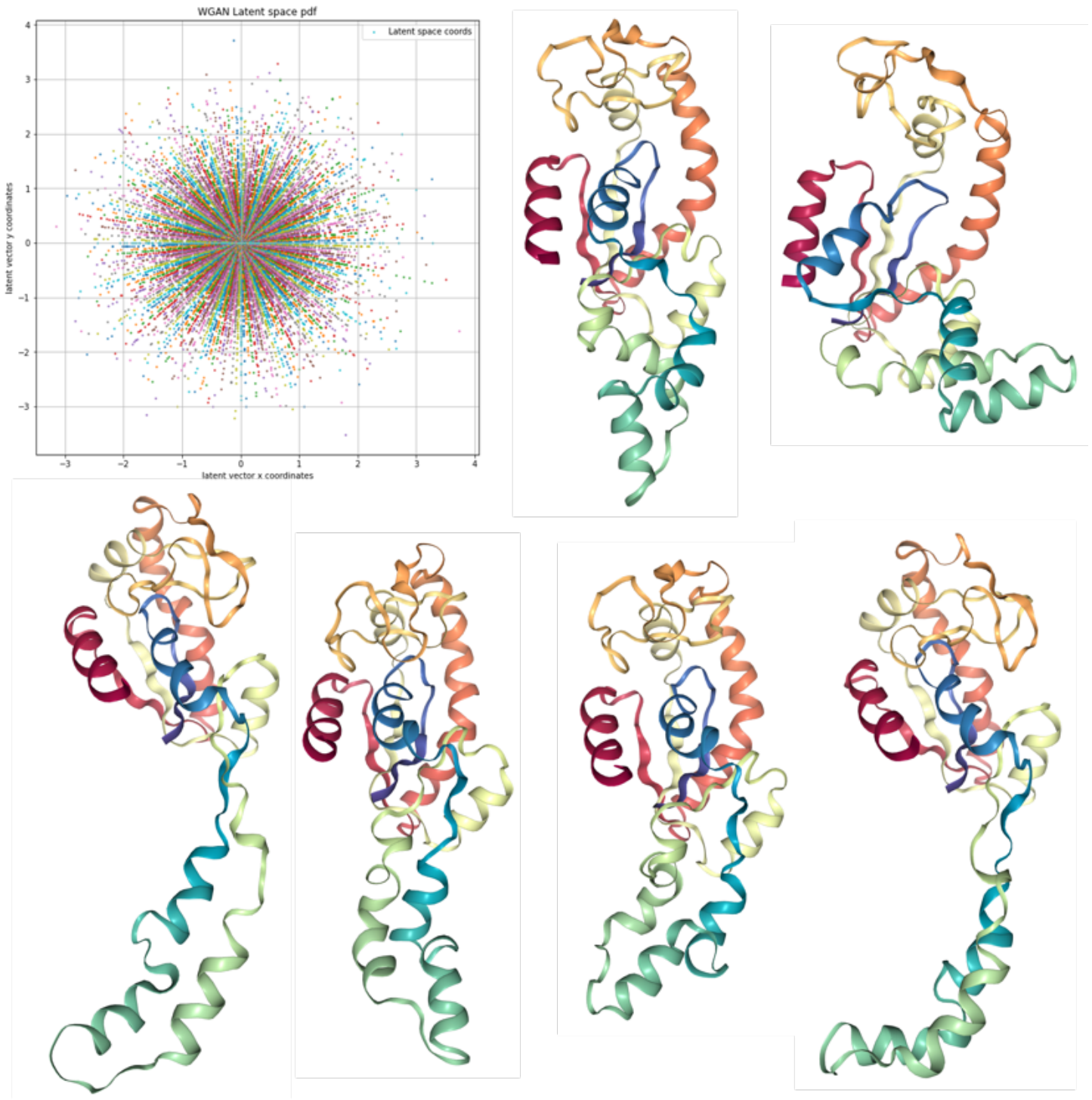

**Figure 9. Boxed panel.** Sampling of the GAN latent space. **Other panels.** Snapshots of AdK conformations reconstructed from basins vectors decoded by the generator from the latent space points shown in the boxed panel.

Our study paves the way for the utilization of these models also in other tasks of biological relevance. For example, *ab initio* protein folding could be achieved by training the generative models with an ensemble of short all-atom MDs that are not long enough to achieve a complete folded state, but sufficient to inform the models on the possible intermediate torsional states of the protein backbone that lead to a complete folding. An important property of neural networks is their capacity to transfer the information learned from one dataset to the training of a different dataset, a process called *transfer learning* [26]. The convolutional nature of the VAE and GAN networks used in this study makes them independent from the number of basins in different proteins. Thus, the *convolutional base* of such networks, trained on the trajectory of a protein that folds on a μs time scale, could be used to initialize or even fix the parameters of a different networks that trains on short dynamics of a protein that folds on a much longer time scale. Finally, it is possible to envision the development of *hybrid* methods combining all-atom with atom-less dynamics, in which the atom-less generative component (Transformer, VAE, or GAN) would provide backbone torsional constraints capable of steering the particle-based component, thus allowing for much longer integration time steps.

In conclusion, we have introduced a strategy for the implementation of different forms of *molecular dynamics without molecules* that may enable the reconstruction of biologically relevant protein states that are difficult to detect with traditional MD computations.

## Conflicts of Interest

The author declares no conflicts of interest.

## Data Repository

All generative methods used in this study were implemented using Keras [23] deep learning libraries running on top of TensorFlow [27]. They are publicly available as Jupyter notebooks at https://github.com/dgattiwsu/MD_without_molecules .

## ORCID

GS: https://orcid.org/0000-0002-9674-261X
LLP: https://orcid.org/0000-0002-3447-1535
AF: https://orcid.org/0000-0002-5102-4294
LS: https://orcid.org/0000-0002-7167-7436
DLG: https://orcid.org/0000-0002-6357-3530